\documentclass{aip-cp}

\usepackage[numbers]{natbib}
\usepackage{rotating}
\usepackage{graphicx}

\begin{document}

\title{CTA in the Context of Searches for Particle Dark Matter - a glimpse}

\author[aff1,aff3]{Jan Conrad}
\eaddress{conrad@fysik.su.se}

\affil[aff1]{Oskar Klein Centre, Physics Department, Stockholm University, SE-10691 Stockholm}
\affil[aff3]{Wallenberg Academy Fellow}

\maketitle

\begin{abstract}
  In this contribution, CTAs potential role in detection of particle dark matter in the context of other detection approaches is briefly discussed for an audience of gamma-ray astronomers. In particular searches for new particles at the large hadron collider and detection of dark matter particles in deep underground detectors are considered. We  will focus on Weakly Interacting Massive Particles (WIMP). Approaches will be compared in terms of (a) robustness of sensitivity predictions, (b) timeline and (c) reach. The estimate of the reach will be model-dependent.  Given our ignorance about the nature of dark matter, and the complementarity of detection techniques even within a given framework (e.g. Supersymmetry), the trivial conclusion is that we might need all approaches and the most sensitive experiments. Our discussion will be somewhat more restrictive in order to be able to be more concrete. With the caveat of incompleteness, under the assumption that the WIMP paradigm describes nature, CTA is more likely to discover multi-TeV WIMP dark matter, whereas for lower masses direct detection and LHC has significantly better prospects. We will illustrate this conclusion with examples from foremost Supersymmetry, but mention effective field theory or simplified models. We comment on a few models predicting high mass WIMPs, in particular 1 TeV higgsino and wino WIMPs, as well as Minimal Dark Matter and  point out the relevance of updated measurements of the anomalous magnetic moment of the muon for CTAs role in searches for Supersymmetry.
\end{abstract}

\section{INTRODUCTION}
The existence of dark matter and its particle nature is established beyond  reasonable doubt, e.g. \cite{Bergstrom:2012fi}. We know that dark matter particles have to be as yet undiscovered with proposed candidates spanning  about 40 orders of magnitude in mass and 50 orders of magnitude in interaction strengths. Experimentalists need more definite predictions to build detectors, which is why experimental approaches have to focus on essentially one particular type of particle at a time.  The particle type that has been the most popular in the last decades, constituting almost a paradigm, is the WIMP (Weakly Interacting Massive Particle). Its popularity has at least two reasons: (a) thermal production of particles, calibrated by light element abundances, predict the relic dark matter abundance of a non-relativistic particle from the (thermally averaged) annihilation cross-section. The very precise measurement of the dark matter density yields an annihilation cross-section that is within factors of few within the cross-section that one would expect from a weakly interacting particle with weak-scale masses, a  numerical coincidence sometimes termed the  ``WIMP miracle''. An non-negligible additional motivation for WIMPs comes from the fact that one of the strongest contenders for physics beyond the standard model, namely supersymmetry (SUSY), provides WIMPs generically, though being proposed for completely different reasons, mainly the hierarchy problem and the unification of gauge couplings. The hierarchy problem, also framed as requiring  naturalness or the absence of fine tuning, prefers WIMP masses below about a $\sim$TeV, which is considerably more constraining than the WIMP coincidence which essentially allows particles from a few GeV to about 100 TeV. As we now know the Higgs mass and have not detected supersymmetric particles yet naturalness is increasingly questioned as guiding principle, see e.g. \cite{Dine:2015xga}, also implying that the scale of supersymmetry (and in particular masses of WIMPs, which are the lightest stable supersymmetric particles) can be  driven to higher values. Gauge unification remains as a motivation, see e.g. split-supersymmetry \cite{ArkaniHamed:2004yi}.\\

\noindent
WIMPs are searched for by three different techniques: at colliders, foremost the LHC, by trying to tag their production in proton-proton collisions, deep underground in ultra-low-background experiments, by trying to detect the scattering of WIMPs of detector material (in jargon called ``direct detection''), or by searching for decay or annihilation products, foremost gamma rays, neutrinos and charged cosmic-rays, from intragalactic, extragalactic or solar system sources  (``indirect detection''). Detection of WIMPs at colliders will strictly speaking require confirmation in one of the other ways to really establish its nature as dark matter.\\

\noindent
CTA, being an Imaging Air Cherenkov Telescope for detection of gamma rays, falls in the category of dark matter indirect detection. The aim of this contribution is to compare CTAs potential to discover WIMP dark matter considering the other ways of detecting WIMPs in terms of (i) timeline, (ii) robustness and (iii) physics reach. As the underlying theory dictates the signatures for (iii) we will have to choose a model, which is difficult even only in the context of supersymmetry, who has a large variety of implementations differing in assumptions and in dimensionality from 5 parameters to more than a hundred. This lack of definiteness has lead to the consideration of effective field theory and lately so called simplified theoretical models to arrive at more model independent conclusions.\\

\noindent
What is attempted here is rather superficial and incomplete, but I believe it would be worthwhile to redo in a more complete fashion. For an example, where a similar approach is taken but more quantitative and in depth and in  the context of SUSY in direct detection prospects and current indirect constraints, see \cite{Baer:2016ucr}, incidentally reaching similar but much more detailed conclusions.

\section{CTA Searches for Particle Dark Matter}
The flux of gamma rays produced in WIMP annihilation is proportional to the integral of the square of the dark matter density over the target solid angle over the line of sight (in jargon referred to as ``J-factor''). Consequently, astrophysical objects with enhanced dark matter density emerge as natural targets to look for dark matter products. For a detailed discussion of the challenges and opportunities of these target, the reader is referred to \cite{Conrad:2015bsa}. Here it suffices to consider two targets: (i) dwarf spheroidal satellites of the milky way  and (ii) the Galactic Center (GC) (or its close vicinity), which are the most relevant targets for CTA for different reasons. For standard non-self interacting cold dark matter unperturbed by baryons the J-factor of the Galactic Center can be very high assuming a cusped DM density profile.  Assuming a cored profile reduces the J-factor (and thereby the expected sensitivity) by about two to three orders of magnitude, without much observational guidance on what would be closer to the truth. Taking cusped J-factors at face value, the GC provides very interesting constraints which directly constrain the above mentioned annihilation cross-section needed to explain the entirety of dark matter observed. Dwarf galaxies on the other hand provide means to estimate the dark matter density by stellar spectroscopy, e.g. \cite{Chiappo:2016xfs}\cite{Geringer-Sameth:2014yza} and references therein. This introduces a statistical uncertainty in the J-factor (as it is at least partly derived from measurements), but also reduces the systematic uncertainty (as it is partly derived from measurements) to arguably a factor two to four (at least for a combined analysis of several dwarfs). Unfortunately, the J-factors for individual dwarfs are lower than the J-factor for the GC by a one to two orders of magnitude, and even in combination probably the reach of CTA using dwarfs as target will be likely within a factor 10 of the target thermal production cross-section. Here we will somewhat optimistically consider CTA GC observations as studied e.g. in \cite{Silverwood:2014yza}\cite{Doro:2012xx}\cite{Carr:2015hta}.

\section{CTA versus other indirect probes}
Except for existing gamma-ray telescopes, foremost the Fermi Large Area Telescope (Fermi-LAT), CTA will have to face the competition of other indirect probes, i.e. charged cosmic rays, neutrinos and (retrieving somewhat less attention) the Cosmic Microwave Background. Indirect searches in charged cosmic rays are dominated by the spectrometer, AMS, on board the international space station (see Schael's contribution in these proceedings), indirect limits from neutrinos are provided by IceCube, e.g.\cite{Aartsen:2014hva} and ANTARES \cite{Adrian-Martinez:2015wey}.\\

\noindent
Predictions for sensitivities or upper limits in other probes than gamma rays are very much subject to similar  systematic uncertainties (foremost ignorance about the dark matter density or systematics dominated backgrounds)  with potentially the added complication of having to model the diffusion and interactions of charged cosmic rays in the galaxy.\\

\begin{figure}[t!]
  \centering\includegraphics[width=0.6\linewidth, bb= 100 0 770 630]{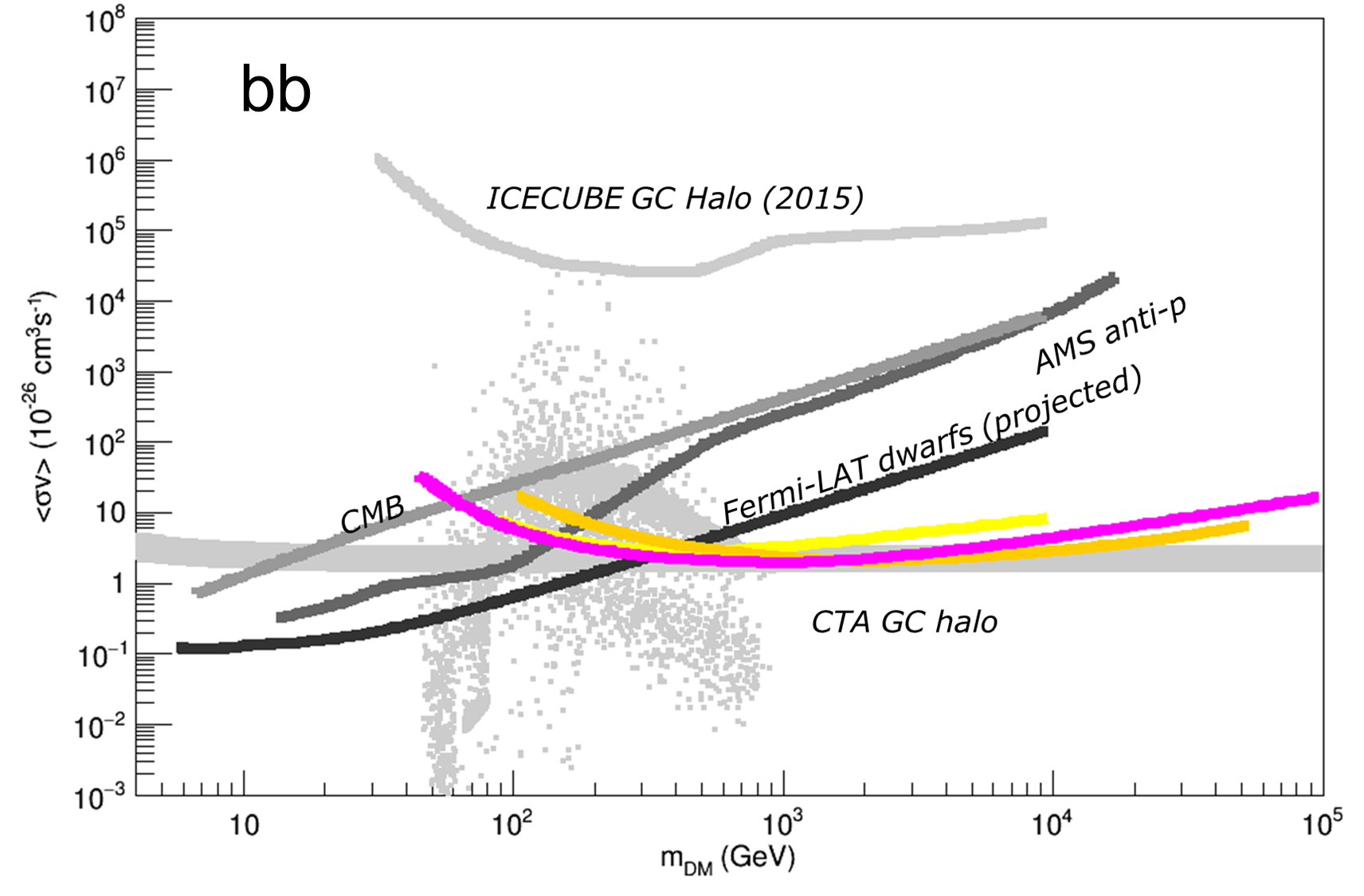}
\caption{Comparison of constraints in thermally averaged annihilation cross-section versus dark matter mass obtained by indirect probes. IceCube GC halo \cite{Aartsen:2014hva}, CMB \cite{Ade:2015xua}, AMS anti-protons \cite{Giesen:2015ufa} and Fermi-LAT dwarfs projection \cite{Charles:2016pgz}. References for CTA projections see text. The gray points illustrate models in pMSSM7.}
\label{fig:indir}
\end{figure}

\noindent
Reach comparison is reasonably model-independent, modulo the assumed annihilation channel. Leptonic channels are generically relatively better for neutrino telescopes/AMS as compared to quark annihilation. The tau-lepton channel, due to its harder spectrum is generically better for high-threshold instruments. An attempt is made to summarize the results in figure \ref{fig:indir} here for the b-quark channel. CTA becomes better than the Fermi-LAT above about 300-500 GeV. At masses of about $\sim$ 1 TeV or higher no other indirect probe can compete with CTA in general. Is there anything else in the future to compete with CTA in terms of gamma-ray telescopes? Novel satellite mission generically target the energy range below the Fermi-LAT and rather address specific issues than providing a large increase in sensitivity. DAMPE launched in 2015 \cite{dampe}, GAMMA-400 e.g \cite{Topchiev:2016whd} and HERD e.g. \cite{Zhang:2014qga} can hope to compete with the Fermi-LAT in searches for line features stemming from dark matter direct annihilation to gamma rays/gamma rays plus boson respectively, due to their excellent energy resolution. On the other hand.  e-Astrogam/PANGU e.g. \cite{Tatischeff:2016ykb} (see also Marco Tavani's contribution to these proceedings) provide excellent angular resolution, thereby providing the potential to resolve the Galactic Center, in order to address the alleged Galactic Center excess, e.g. \cite{Calore:2015hjq} and references therein. Apart from this particular aspects, it seems it will take of the order of 10 years at least to compete with results from the Fermi-LAT.

\section{CTA versus production at the Large Hadron Collider}
At the Large Hadron Collider (LHC)  potential WIMPs can be produced in proton-proton collisions and are searched for either in decay chains for SUSY particles, e.g. \cite{Rammensee:2015mea} , more generically WIMP final states with missing transverse energy or in the so called mono-X signature, where X is a jet,leptonic or photon signature due to initial state radiation and the final state is the (invisible) dark matter particle(s), see e.g. \cite{Meirose:2016pxn}.  It is to be reiterated, but even if the newly discovered particle has the required properties (stability is obviously hard to assess), it would be necessary to be identified by astrophysical observations.\\ 

\noindent
To our knowledge most predictions for the LHC have been made for an integrated  luminosity of 300 fb$^{-1}$, and this assumption is going to be reached at the end of Run III, i.e. 2023. Sensitivity predictions are based on detailed Monte Carlo and often background estimates are data-driven and can therefore be considered robust and accurate at least from the experimental side. Claimed discoveries will be solid.\\

\noindent
For the reach pMSSM19 has been considered, e.g. \cite{Cahill-Rowley:2014boa}. Effective field theory is more complicated for comparison with LHC due to its potentially restricted validity, see e.g.\cite{Buchmueller:2013dya}. Simplified models then provide an alternative, but studies are still scarce for the comparison between LHC and indirect detection.\\

\noindent
We take the opportunity to give a few general remarks with respect to global fits and Supersymmetry:  apart from the assumed implementation of supersymmetric model, SUSY global fits differ in a variety of ways, foremost which observables are included and what statistical methodology is employed. While some groups perform Bayesian or frequentist global fits, i.e. statistically favoured regions can be defined according to some statistical measure (at least under the non-trivial assumption that the likelihood is sufficiently well sampled), others present a large set of models and essentially count models surviving under a certain conditions, which (at least) very difficult statistical interpretation. This is the case for the pMSSM19 comparison between LHC and CTA \cite{Cahill-Rowley:2014boa}, that we use here.\\

\noindent
One particularly interesting  observation in the context of CTA is the anomalous magnetic moment of the muon, $\delta (g-2)_\mu$, which shows a $\sim$ 2.7 $\sigma$ deviation from the expectation from the standard model \cite{Bennett:2006fi}. If implemented in SUSY global fits, this value disfavours high WIMP masses (e.g. for the pMSSM $\gtrsim$ 1 TeV, see \cite{Bagnaschi:2015eha}, which however does not consider indirect detection) and figure \ref{fig:bagnaschi}.  Updates of  $\delta (g-2)_\mu$ are hopefully coming already in 2017 \cite{Gray:2015qna} and will be crucial for CTAs role in searches for Supersymmetry.\\

\begin{figure}[t!]
  \centering\includegraphics[width=0.6\linewidth, bb= 0 0 650 330 ]{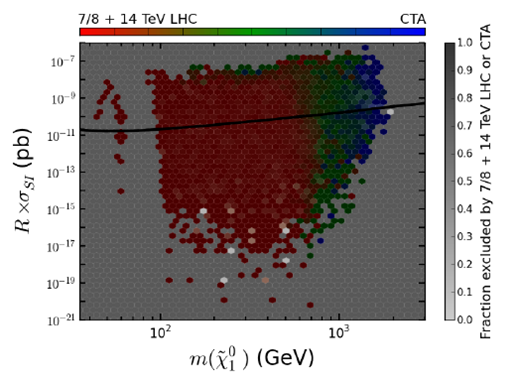}
\caption{Model point ``densities'' in a plane of spin-independent scattering cross-section (rescaled for the case were the dark matter is only partly provided by WIMPs) versus (neutralino-)WIMP mass. Colour-scale indicates reach of LHC and CTA. The solid line indicates reach of 1 tonne scale direct detection experiments. Note that XENONnT/LZ will probe cross-sections down to $10^{-12}$pb.}
\label{fig:cahill-rawley}
\end{figure}

\noindent
The work of \cite{Cahill-Rowley:2014boa}, where $\delta (g-2)_\mu$  is not included and where models are not required to provide the entirety of dark matter may serve as an illustration for comparing the reach of CTA and LHC. In figure \ref{fig:cahill-rawley} one can see how LHC sensitivity is restricted to below around one TeV. Straw-men effective field theory models, like the one presented in \cite{Arrenberg:2013rzp} confirm this conclusion qualitatively. A full likelihood inference including LHC Run III prospects (but excluding $\delta (g-2)_\mu$ for the time being) to understand whether CTAs unique capabilities reach pMSSM models that are not already disfavoured would be interesting.

\section{CTA versus Direct Detection}
In the mass range of above 10 GeV or so, the most sensitive direct detection detectors are time projection chambers (TPCs) made of xenon. The combination of liquid xenon and gaseous xenon allows both a light signal and charge signal to be read out, which in turn allows background discrimination. The reach of xenon TPCs  has evolved quickly: in the last decade  the sensitivity to the WIMP nucleus (spin-independent) scattering has improved by almost two orders of magnitude, see  e.g. \cite{Baudis:2014naa}. This development was lead by the XENON programme, e.g. \cite{Aprile:2012nq}\cite{Aprile:2016swn} at LNGS, recently superseded in sensitivity by the LUX xenon TPC at the Sanford Underground Facility in South Dakota \cite{Akerib:2016vxi} and the Chinese xenon TPC,  PandaX \cite{Tan:2016zwf}.\\

\noindent
Near-future experiments will improve sensitivities by another two orders of magnitude: the XENON1T experiment \cite{Aprile:2015uzo} which is about to start science operations and its upgrade XENONnT, envisaged to start operations in 2018 as well as the successor of LUX, the LUX-ZEPLIN (LZ) experiment \cite{Akerib:2015cja}. This implies that roughly by 2021 scattering cross-sections of $\sigma_{scattering} \approx 10^{-12}$ pb will be in reach.\\

\noindent
Uncertainties in sensitivity predictions are mainly due the local WIMP density (factor $\sim$2 uncertainty entering linearly into cross-section constraints \cite{Read:2014qva}) and velocity distribution (important for small masses, but probably small for masses $>$ 20 GeV, see e.g. the recent results including baryons \cite{Sloane:2016kyi} \cite{Bozorgnia:2016ogo}) as well as calculating the response of nuclei to WIMP interaction (e.g. \cite{Gresham:2014vja} indicates a small effect for xenon TPCs)\footnote{the nucleon-dark matter interaction assumptions can introduce a significant theoretical bias, e.g.  \cite{Catena:2014hla}. The standard assumption is that  the  dark  matter-nucleon  scattering is  momentum  and  velocity  independent (motivated by small halo velocities) corresponding to spin-independent interaction and spin-dependent interaction.}. Above masses of 20-50 GeV, which is the relevant for comparison with CTA, a signal can probably be considered a smoking gun and predicted sensitivies are rather robust.\\

\begin{figure}[t!]
  \centering\includegraphics[width=0.7\linewidth, bb= -20 0 570 330 ]{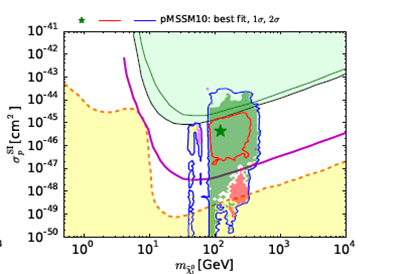}
  \caption{A global fit within pMSSM10 presented by  \cite{Bagnaschi:2015eha}. This fit includes LHC 8 TeV and $\delta (g-2)_\mu$ (among other observables), which explains the disvavoured high mass WIMP region. Results are presented in a plane spin-independent scattering cross-section versus neutralino mass. The yellow shaded region corresponds to the parameter space where neutrino background can not be discriminated from WIMP signal. The magenta line is approximately the reach of XENONnT/LZ. The green star is the best fit, the solid contours correspond to 1$\sigma$, 2$\sigma$ respectively. It would be interesting to redo this fit for pMSSM removing the $\delta (g-2)_\mu$ constraints and including LHC 14TeV results/prospects.}
\label{fig:bagnaschi}
\end{figure}

\noindent
As an example, for the  pMSSM10 global fit of \cite{Bagnaschi:2015eha}, future direct detection experiments would only leave already disfavoured (at $>$ 2 sigma level) SUSY space unprobed, see figure \ref{fig:bagnaschi}. Note that these scans do include the above mentioned $\delta (g-2)_\mu$. A global fit in pMSSM19 without this observable is presented in \cite{Roszkowski:2014iqa}. One particularly interesting candidate is a $\sim$ 1 TeV higgsino-like WIMP. For this candidate, direct detection is likely to be able to probe the most favoured parameter space (see e.g. \cite{Fowlie:2013oua}),  but CTA can potentially probe two-three orders of magnitude deeper (in terms of scattering) than XENONnT/LZ.  At higher masses (see next section) the situation looks more favourable for indirect detection as for even higher mass WIMPs resonance effects can increase the annihilation cross-section. Simple EFT calculations  \cite{Arrenberg:2013rzp} indicate as well that CTA will be more sensitive than direct detection for multi-TeV WIMP masses.

\begin{figure}[t!]
  \centering\includegraphics[width=0.7\linewidth, bb= -20 0 380 330]{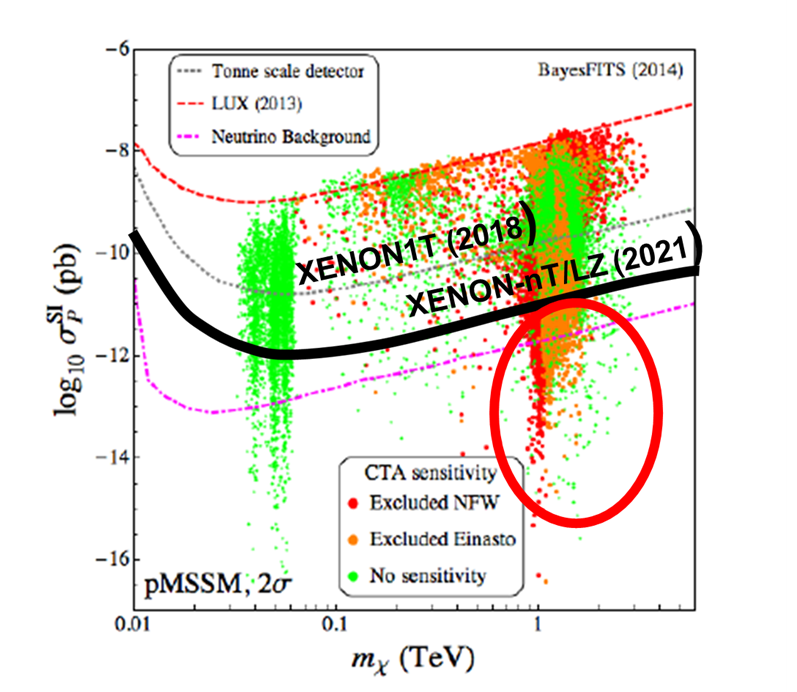}
  \caption{Global fit to pMSSM19 presented by \cite{Roszkowski:2014iqa}. Shown are points that are consistent with the best fit on 2$\sigma$ level (including relic density, but not $\delta (g-2)_\mu$), in the plane of spin-independent scattering cross-section versus neutralino mass. The 1 TeV region consists of  mostly Higgsino like WIMPs. The approximate projection for XENONnT/LZ was added here by the author. The figure illustrates how CTA can probe models out of reach of next generation direct detection experiments. CTA also has a fair chance to make complemenary measurements, should direct detection experiments discover the Higgsino WIMP.}
\label{fig:roskowski}
\end{figure}  

\section{High mass WIMPs}
The most model independent conclusion we can draw so far is that multi-TeV is the region where CTA will have the highest potential of first discovery. The question arises whether there are well motivated models that provide WIMPs of that mass. Within pMSSM (also in Split SUSY), one of the possible candidates beyond TeV is Wino Dark Matter (WinoDM). The phenomenology of WinoDM is analogous to a class of DM models that augment the Standard Model with fermionic multiplet states, WinoDM corrsponds to the triplet state. For indirect detection the relevance of these class of models comes from the fact that their annihilation cross-section can be enhanced by a non-relativistic resonance effect (Sommerfeld enhancement). Requiring these models to provide dark matter due to thermal production fixes the expected mass of the WIMP (in WinoDM to about 3 TeV) and is essentially excluded  by Fermi-LAT dwarf observations \cite{Ackermann:2015zua} and HESS observations of the Galactic Centre \cite{HESS::2016jja}. However, this is only true for a cusped profile of the dark matter distribution in the galactic center. In fact, while higgsino WIMPs will fall out of reach for CTA in case of cored DM profiles, WinoDM will not be excluded anymore, but within reach for CTA. Also observations of dwarf galaxy spheroidals will be relevant in this case, which will open the possibility to obtain relevant constraints with empirically determined J-factors.\\

\begin{figure}[t!]
  \centering\includegraphics[width=0.6\linewidth, bb = 50 0 450 450]{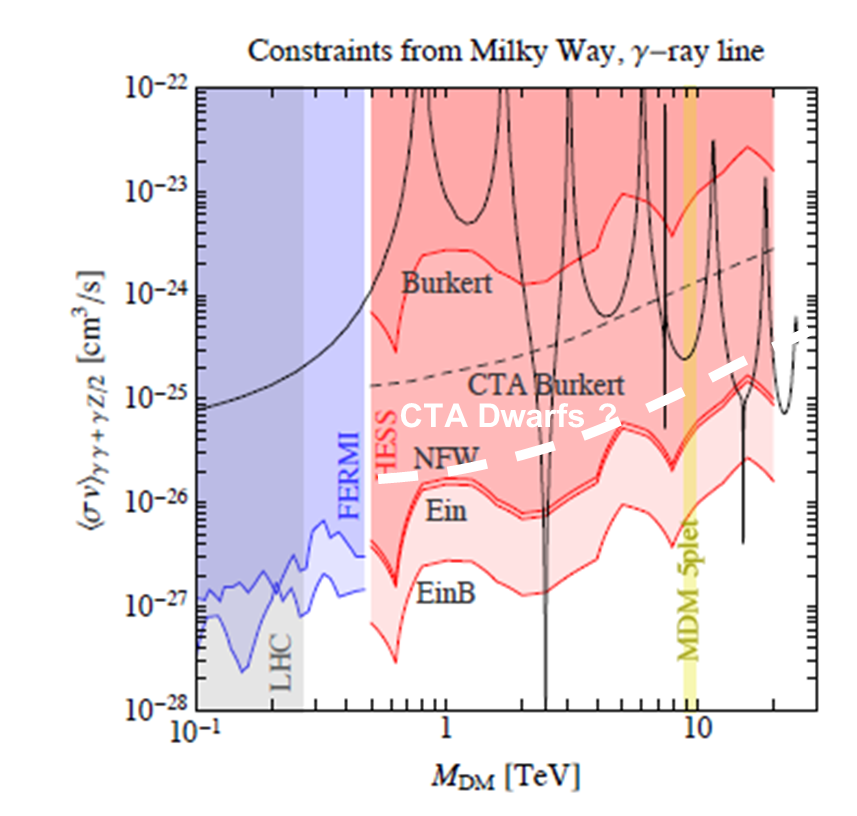}
\caption{Gamma-ray constraints compared to predictions for MDM in a plane of line annihilation cross-section versus mass \cite{Cirelli:2015bda}. Shown is the MDM quintuplet providing dark matter and constraints of HESS, Fermi-LAT and predictions for CTA under different assumptions on the dark matter density profile. HESS searches exclude MDM if the profile is cusped. CTA searches, however, can exclude the predictions even if the profile is cored (Burkert). I include here my estimate of a combined CTA dwarf analysis maximal reach for illustration.}
\label{fig:MDM}
\end{figure}

\noindent
This line of reasoning probably holds for a more general class of multi-TeV WIMP models (at least if they annihilate generically into similar states). Another concrete example is Minimial Dark Matter (MDM) \cite{Cirelli:2005uq}, a fermionic quintuplett state, where the WIMP has a predicted mass of about 10 TeV, see figure \ref{fig:MDM} for results regarding MDM in the context of indirect detection with gamma rays.\\

\noindent
These models are clearly out of reach for LHC, for direct detection, the prospects are poorly studied. MDM WIMPs might be close to the reach for XENONnT/LZ scale experiments \cite{Cirellib}.

\section{Discussion}
For an experimentalist, discussions like the one presented here are performed only when asked to prioritize resources. Even if restricting ourselves to only consider a Supersymmetric WIMP, no single  detection technique will cover the full parameter space. More detailed comparison of prospects are largely depending on the actual version of Supersymmetry considered. Studies in EFT and simplified models, which partly overcome this problem are still relatively scarce. Given the timeline and relative robustness of prospects for LHC and direct detection in the next five to seven years, however, it seems likely that generically WIMPs, and specifically generically SUSY WIMPs, will be either discovered or excluded by these techniques if their mass is below about a few TeV. CTA should prioritize accordingly, if discovery is the aim. Finally, I mention that updated measurements of $\delta (g-2)_\mu$ (potentially forthcoming next year) will be very interesting for CTA in the context of search for supersymmetry.


\section{ACKNOWLEDGMENTS}
I had a number of informal discussion for this talk, in particular I thank: E. Bagnaschi, C. Balazs, O. Buchmueller, R. Catena, M. Cirelli, J. Edsj\"o and L. Roszkowski.


\bibliographystyle{natbib}%

\end{document}